\documentclass{emulateapj}

\usepackage [english]{babel}
\usepackage [english = american]{csquotes}
\MakeOuterQuote{"}

\usepackage{xspace}

\newfont{\tensy}{cmsy10}
\newcommand{\chemical}[1]{{$\fontdimen16\tensy=3.0pt \fontdimen17\tensy=3.0pt
\mathrm{#1}$}}
\newcommand{\Ti} {\chemical{^{44}Ti}\xspace}
\newcommand{\Sc} {\chemical{^{44}Sc}\xspace}

\newcommand{\G} {G1.9+0.3\xspace}
\newcommand{\NUSTAR} {\textit{NuSTAR}\xspace}
\newcommand{\NUSTARs} {\textit{NuSTAR's}\xspace}
\newcommand{\CHANDRA} {\textit{Chandra}\xspace}
\newcommand{\COMPTEL} {\textit{Comptel}\xspace}
\newcommand{\INTEGRAL} {\textit{INTEGRAL}\xspace}
\newcommand{\NUSTARDAS} {\textit{NuSTARDAS}\xspace}

\newcommand{\chis} {$\chi^2$\xspace}

\newcommand{\NH} {N$_{\rm H}$\xspace}
\newcommand{\cmnt} {$\mathrm{cm}^{-2}$\xspace}
\newcommand{\Flux} {$\,\mathrm{ph}\,\mathrm{cm}^{-2}\,\mathrm{s}^{-1}$\xspace}

\catcode`\@=11
\def\gapprox{\mathrel{\mathpalette\@versim>}}
\def\lapprox{\mathrel{\mathpalette\@versim<}}
\def\@versim#1#2{\lower2.9truept\vbox{\baselineskip0pt\lineskip0.5truept
    \ialign{$\m@th#1\hfil##\hfil$\crcr#2\crcr\sim\crcr}}}
\catcode`\@=12

\shorttitle{The Hard X-Ray View of the Young Supernova Remnant G1.9+0.3}
\shortauthors{Andreas Zoglauer et al.}

\begin{document}

\submitted{Accepted ApJ}

\title{The Hard X-Ray View of the Young Supernova Remnant G1.9+0.3}
 

\author{Andreas Zoglauer\altaffilmark{1}, Stephen P. Reynolds\altaffilmark{2}, Hongjun An\altaffilmark{3}, Steven E. Boggs\altaffilmark{1}, Finn E. Christensen\altaffilmark{4}, William W. Craig\altaffilmark{1,5}, Chris L. Fryer\altaffilmark{6}, Brian W. Grefenstette\altaffilmark{7}, Fiona A. Harrison\altaffilmark{7}, Charles J. Hailey\altaffilmark{8}, Roman A. Krivonos\altaffilmark{1}, Kristin K. Madsen\altaffilmark{7}, Hiromasa Miyasaka\altaffilmark{7}, Daniel Stern\altaffilmark{9}, William W. Zhang\altaffilmark{10}}

\email{Corresponding author: Andreas Zoglauer, zog@ssl.berkeley.edu}
\altaffiltext{1}{Space Sciences Laboratory, University of California, Berkeley, CA 94720, USA}
\altaffiltext{2}{Physics Department, North Carolina State University, Raleigh, NC 27695, USA}
\altaffiltext{3}{Department of Physics, McGill University, Montreal, Quebec, H3A 2T8, Canada }
\altaffiltext{4}{DTU Space, National Space Institute, Technical University of Denmark, Elektrovej 327, DK-2800 Lyngby, Denmark }
\altaffiltext{5}{Lawrence Livermore National Laboratory, Livermore, CA 94550, USA }
\altaffiltext{6}{CCS-2, Los Alamos National Laboratory, Los Alamos, NM 87545, USA }
\altaffiltext{7}{Cahill Center for Astronomy and Astrophysics, California Institute of Technology, Pasadena, CA 91125, USA }
\altaffiltext{8}{Columbia Astrophysics Laboratory, Columbia University, New York, NY 10027, USA }
\altaffiltext{9}{Jet Propulsion Laboratory, California Institute of Technology, Pasadena, CA 91109, USA }
\altaffiltext{10}{Goddard Space Flight Center, Greenbelt, MD 20771, USA }







\begin{abstract}

\NUSTAR observed G1.9+0.3, the youngest known
supernova remnant in the Milky Way, for 350~ks and detected emission up to $\sim$30~keV.  The
remnant's X-ray morphology does not change significantly across the energy
range from 3 to 20~keV.  A combined fit between \NUSTAR and \CHANDRA
shows that the spectrum steepens with energy.  The spectral
shape can be well fitted with synchrotron emission from a power-law
electron energy distribution with an exponential cutoff with no
additional features.  It can also be described by a purely
phenomenological model such as a broken power-law or a power-law with
an exponential cutoff, though these descriptions lack physical
motivation.  Using a fixed radio flux at 1~GHz of 1.17~Jy for the
synchrotron model, we get a column density of N$_{\rm H}$~=~$(7.23\pm0.07)
\times 10^{22}$~cm$^{-2}$, a spectral index of $\alpha=0.633\pm0.003$,
and a roll-off frequency of $\nu_{\rm rolloff}=(3.07\pm0.18) \times 10^{17}$~Hz.
This can be explained by particle acceleration, to a maximum energy
set by the finite remnant age, in a magnetic field of about 10 $\mu$G,
for which our roll-off implies a maximum energy of about 100 TeV for
both electrons and ions.  Much higher magnetic-field strengths would
produce an electron spectrum that was cut off by radiative losses,
giving a much higher roll-off frequency that is independent of
magnetic-field strength.  In this case, ions could be accelerated to
much higher energies.  A search for $^{44}$Ti emission in the 67.9~keV line
results in an upper limit of $1.5 \times 10^{-5}$ $\,\mathrm{ph}\,\mathrm{cm}^{-2}\,\mathrm{s}^{-1}$ assuming a line
width of 4.0~keV (1~sigma).
\end{abstract}





\section{Introduction}

Both core-collapse and Type Ia
supernova remnants (SNRs) are primarily known as radio objects which
exhibit synchrotron radiation from relativistic electrons with
energies in the GeV range.  However, in a few young remnants with high
shock velocities, this synchrotron spectrum is observed to continue
into the X-ray regime. This requires electrons in the TeV range 
\citep[for a review see][]{Reynolds2008a}.

While the typically featureless power-law spectrum in the radio band
contains relatively little detailed information on the processes by
which shocks accelerate particles, the X-ray spectrum shows a
high-energy cutoff, giving direct information on the
maximum energies to which particles are being accelerated.  The
detailed spectral shape and morphology of synchrotron X-ray emission
provide powerful constraints on quantities such as the diffusion
coefficient and the mean magnetic-field strength.  Most models of
particle acceleration do not predict a sharp cutoff in the particle
distribution.  Instead, exponential cutoffs in energy
\citep[e.g.][]{Drury1991} or in energy squared \citep{Zira2007} have
been proposed, leading to synchrotron spectra dropping no faster than
exponentially in photon energy.  Such spectra can be described by a
characteristic roll-off photon energy corresponding to an electron
energy characterizing the cutoff.

The maximum energy to which SNR shocks can accelerate particles is an
important question in understanding the origin of cosmic rays, and one
accessible to study through observations of the X-ray synchrotron
spectra of young remnants.  The finite time available for particle
acceleration will limit the maximum energy to the same value for both
electrons and ions (the ``age-limited'' case).  However, the electron
distribution may cut off at a lower energy due to radiative losses
(``loss-limited'' acceleration), but since radiative losses are
negligible for ions, their distribution could continue to a much
higher cutoff energy.  Thus, understanding which mechanism is
responsible for the cutoff in an observed spectrum can provide
indirect information on ion acceleration as well.

While a dozen or so young Galactic SNRs show evidence for X-ray
synchrotron emission alongside much stronger thermal emission, only a
handful exhibit X-ray spectra dominated by synchrotron emission.  This
includes the youngest supernova remnant in our Galaxy, \G
\citep{Reynolds2008b}.  This object has the smallest angular size of
any Galactic remnant (about $100''$ in diameter), the highest shock
velocities (from both expansion proper motions and Doppler shifts of
lines from isolated regions of thermal emission, about 14,000 km
s$^{-1}$), and one of the highest roll-off energies observed
($h\nu_{\rm rolloff} \sim 2.2$ keV)
\citep{Reynolds2009,Borkowski2010,Carlton2011}.  The currently
observed expansion rate, along with simple one-dimensional hydrodynamic models,
suggests a deceleration, with shock radius $R$ varying with time $t$ 
since the explosion as
$R \propto t^{0.7}$, and giving an age $t \sim 110$ yr
\citep{Carlton2011}.

Several arguments suggest, but do not compel, a Type Ia origin of \G
\citep[for details see][]{Reynolds2008b, Borkowski2013a}: the high
velocities more than 100 years after the explosion; the absence of any
central pulsar-wind nebula (though thermal emission from a neutron
star would be too highly absorbed to be detectable); the bi-symmetric
X-ray morphology, analogous to SN 1006; and substantial thermal
emission from Fe.  The high absorption (\NH $\sim 5 \times 10^{22}$
\cmnt; \citealt{Reynolds2009}) and substantial distance (estimated at
8.5 kpc) mean that only a relatively small range of photon energies
is accessible to \CHANDRA.  The behavior of the spectrum over a
broader energy range, as accessible with \NUSTAR, is of considerable interest for modeling the
process of shock acceleration.  For example, the difference between
exponential cut-offs in photon energy and in the square root of photon
energy is difficult to discern over the available \CHANDRA bandpass
of 1.5 -- 7 keV, where high absorption provides the lower limit.

In addition, \G is of interest for another important reason: it shows
evidence for the presence of \Ti in the explosion, through an
inner-shell transition at 4.1 keV in \Sc, to which \Ti decays by
electron capture \citep{Borkowski2010}.  The estimated mass in \Ti,
about $1 \times 10^{-5}\ M_\odot$ \citep{Borkowski2010,Borkowski2013b},
implies a flux in the 68 and 78~keV nuclear de-excitation lines of about
$5 \times 10^{-6}$ \Flux.  \G is only the third SNR, after Cas A
\citep{Iyudin1994} and SN 1987A \citep{Grebenev2012}, to show evidence
for radioactive titanium.  If \G resulted from a thermonuclear
explosion, this mass in \Ti provides an important constraint on
models, since it may require substantial asymmetry in the explosion
\citep[e.g.][]{Maeda2010}.  However, the 4.1 keV line is very broad,
and the uncertainties in the amount of \Ti are
considerable. Independent constraints on the flux of \Ti and
correspondingly better constraints on the mass are important goals for
observations with \NUSTAR.

The paper is structured as follows: \S \ref{sec:obs} describes
the \NUSTAR observations, \S \ref{sec:morphology} analyses the
morphology of the SNR, \S \ref{sec:spectral} determines and
interprets the spectral properties of the supernova remnant, and
\S \ref{sec:ti} gives upper limits for the \Ti emission and
yield.

\section{Observations}
\label{sec:obs}

\begin{deluxetable*}{cccc}
\tabletypesize{\scriptsize}
\tablecaption{List of \NUSTAR observations of G1.9+0.3\label{ObsList} in 2013}
\tablewidth{0pt}
\tablehead{
\colhead{ID} & \colhead{Start date [Day - Time]}  & \colhead{Stop date [Day - Time]} & \colhead{Exposure [ks]}
}
\startdata
40001015003 & 189 - 17:25 & 192 - 09:15 & 86 \\
40001015005 & 195 - 02:45 & 197 - 22:05 & 122 \\
40001015007 & 208 - 20:25 & 212 - 01:45 & 145 \\
\enddata
\end{deluxetable*}

The \NUSTAR telescope \citep[see][for a description]{Harrison2013} observed \G in July and August 2013 with an effective exposure time of 353~ksec.
A full list of the observations is provided in Table \ref{ObsList}.
\NUSTAR consists of two co-aligned telescope modules with corresponding focal plane modules termed FPMA and FPMB.
Both operate in the energy range from 3 to 79~keV.
For the analysis, the data from both focal plane modules and all three observation periods were used. 
All data were reduced using the tools included in HEASoft version 6.14 which includes \NUSTARDAS, the \NUSTAR Data Analysis Software (version 1.3.1 with \NUSTAR CALDB version 20131223), as well as custom developed analysis tools based on ROOT \citep{Brun1997}.
During some of the observations the default event selection resulted in an increased background flux while \NUSTAR was close to
the South Atlantic Anomaly (SAA). 
Therefore, we instead used the "optimized SAA cut" to eliminate those time intervals. 
This reduced the effective observation time by roughly 1\%.

\begin{figure*}[ht]
\includegraphics[width=\textwidth]{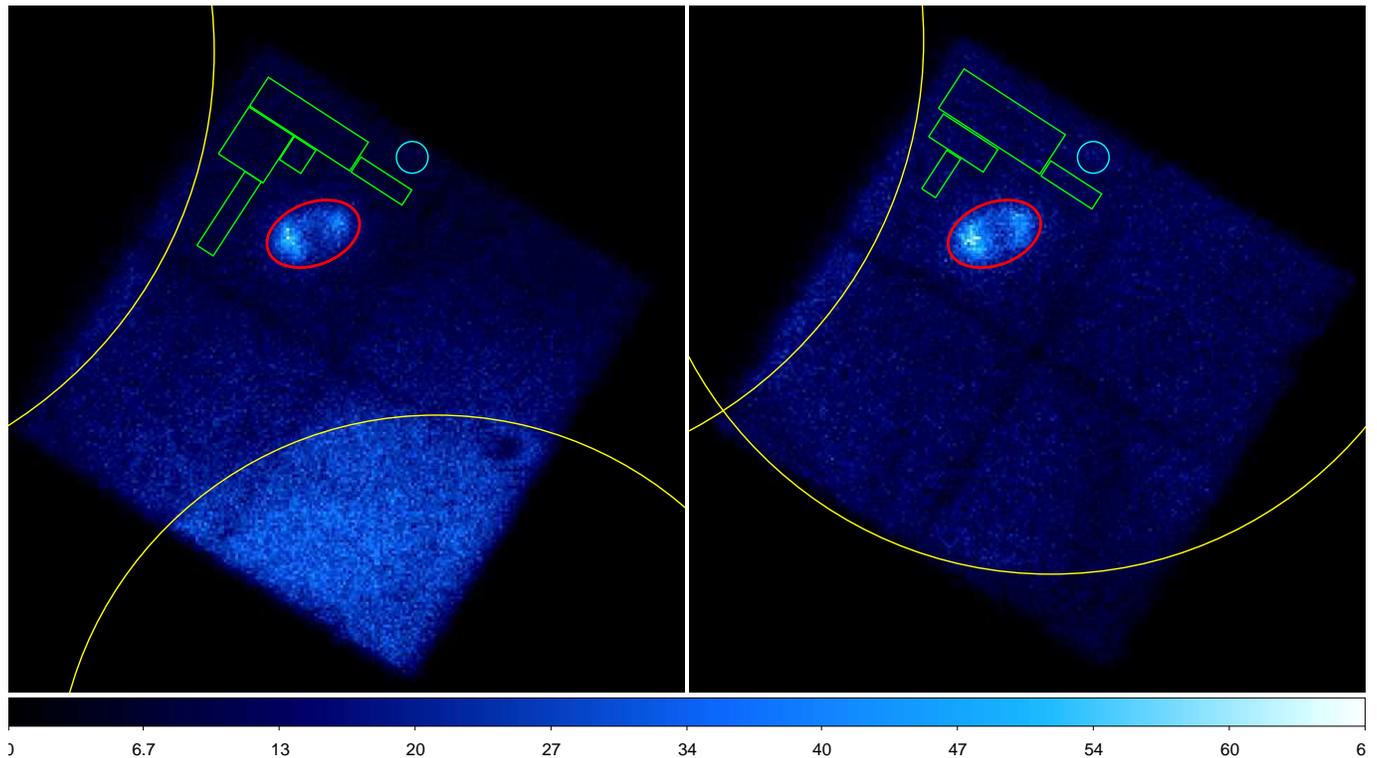}
\caption{Images of \G (not deconvolved) from observation 40001015007 for focal plane module A (left) and B (right). The remnant is located inside the red ellipses. The background extraction region is the combination of the green boxes. The small cyan circle shows the position of a weak source which is only clearly visible when all observations are combined. Within the large yellow circles the background is increased due to stray light.
\label{Fig:Regions}}
\end{figure*}

Figure \ref{Fig:Regions} shows a simple (not deconvolved) image of the longest observation (40001015007, 145~ksec) for both modules. 
The supernova remnant itself is centered on detector zero (top most detector in the image) on both modules. 
Its extraction region for later spectral analysis is marked by red ellipses.
In addition, a weak source at the edge of the detector (white circle) --- which is only clearly visible when all data is combined --- and several "zero-bounce" stray light sources are visible.
For \NUSTAR, zero-bounce stray light sources, i.e.~sources where x-rays hit the detector without impinging on the optics \citep[for details see][]{Wik2014}, manifest themselves by increased emission within large circular regions, shown in yellow in Figure \ref{Fig:Regions}.
On FPMA (left), at least two stray light sources are visible, and the two strongest are marked.
On FPMB (right), two stray light sources are visible. 
While the stray light sources do not intersect the remnant on FPMA, one stray light source completely covers the remnant on FPMB, and consequently increases the background for the source. 
The green rectangles show the background extraction regions. 
They were chosen to maximize the collected background data from the same detector that observed the SNR, while avoiding the wings of the point spread functions of the SNR emission, detector boundaries, stray light sources, as well as the additional weak source at the detector edge.

\section{Morphology} 
\label{sec:morphology}

\begin{figure}[ht]
\includegraphics[width=\columnwidth]{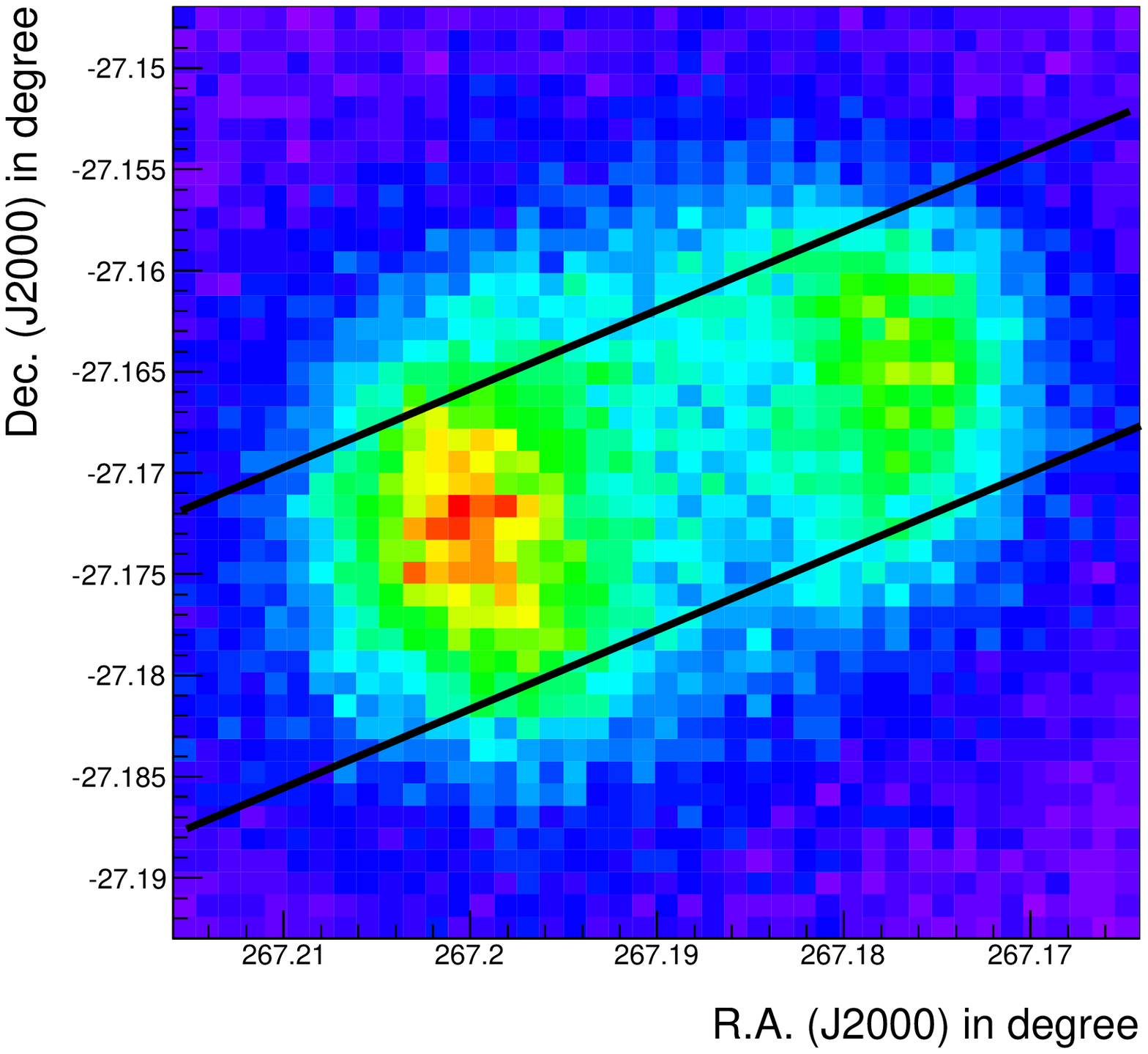}
\caption{Image (not smoothed or deconvolved) of all observed data between 3 and 20~keV. The region between the black lines has been used to generate the profile in Figure \ref{Fig:Profile}.
\label{Fig:AllBackprojected}}
\end{figure}

Figure \ref{Fig:AllBackprojected} shows a simple (not deconvolved) image of \G. 
The data from all three observations and both focal plane modules have been combined. 
Although the remnant shows emission up to 30-40~keV (see next section), above 20~keV the statistics are not sufficient to generate a good image.
Therefore, only photons between 3 and 20~keV have been used to generate the image in Figure \ref{Fig:AllBackprojected}.
The bilaterally symmetric shape of the remnant, first detected with \CHANDRA \citep{Reynolds2008b}, is clearly visible with enhanced X-ray emission from the south-east and north-west corners.
However, due to \NUSTARs point spread function, the central low-emission region, which is clearly visible in \CHANDRA images, as well as the north and south ridge are not obviously visible. 
To retrieve those morphological features we apply image deconvolution techniques in Section \ref{sec:decon}.

\subsection{One-dimensional profile of the remnant}

With \NUSTAR's wide energy band we can now explore whether the X-ray emitting regions are energy dependent.
While we do not have sufficient statistics for meaningful full two-dimensional comparisons of the low-energy and the high-energy emission, we can compare the low and high-energy emission in a one-dimensional profile through the remnant, which covers everything but the north and south ridge.

\begin{figure}[ht]
\vspace{0.5cm}
\includegraphics[width=\columnwidth]{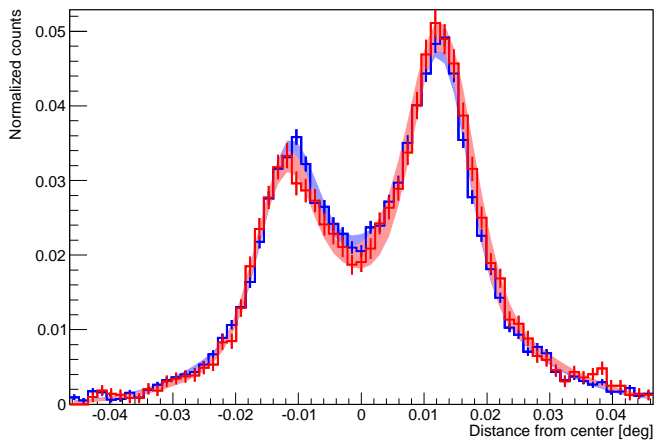}
\caption{\NUSTAR surface brightness profile of \G from the south-east to the north-west for photons in the range 3-8 keV (blue) and 8-20 keV (red). 
The blue and red bands represent the 90\% confidence range of a fit with three Gaussians.
The red high-energy band overplots the blue low-energy band.
Given the angular resolution of \NUSTAR, the two profiles cannot be distinguished with current image statistics.
\label{Fig:Profile}}
\end{figure}

Figure \ref{Fig:Profile} shows the profile of the remnant along a slice from the south-east to the north-west corner for two energy ranges, 3-8~keV (blue) and 8-20~keV (red). 
The profiles have been background subtracted and then normalized to the same area below the curve. 
Both profiles have been fitted by three Gaussians. 
The 90\% confidence bands around those fits have been determined taking into account the statistical uncertainty of each bin. 
In the figure, the red high-energy band overplots the blue low-energy band. 
Both bands overlap over the whole range of the plot.
Therefore, within the given angular resolution and statistics, the low-energy emission in the band from 3-8~keV cannot be distinguished from the 8-20~keV emission.

\subsection{Deconvolved images}
\label{sec:decon}

Another key question is how the \NUSTAR image compares to the \CHANDRA image. 
Due to the lower angular resolution of \NUSTAR (half-power diameter of 58$''$, FWHM of 18$''$), a direct comparison is difficult. However, since the point-spread function (PSF) has a narrow core, it is possible to significantly sharpen the image using deconvolution techniques.

Five main components contribute to the observed \NUSTAR image of \G:
\begin{itemize}
\item source photons focused through the optics, including \G source photons as well as focused cosmic and Galactic diffuse X-rays;
\item "ghost rays" --- source photons passing through the optics with only a single scatter in the optics;
\item stray light from sources passing through the narrow solid angle between the optics and the aperture stop;
\item aperture background --- diffuse X-ray background passing through the narrow solid angle between optics and aperture stop; and
\item internal background from activation and photons leaking through the shield.
\end{itemize}
For more details on the \NUSTAR background see \cite{Wik2014}.

For the purposes of image reconstruction, the internal background is simply a flat offset. 
The aperture background is low compared to the remnant, with sufficiently little spatial variation over the remnant so that it does not need to be considered as a separate component in the image deconvolution.
In addition, our observations contain no clearly visible ghost rays or stray light coincident with the remnant for FPMA, and for FPMB the stray light is reasonably flat across the remnant.
Therefore, for the image deconvolution we simply consider the source photons plus a flat background component.

Image reconstruction for X-ray (and gamma-ray) telescopes consists of two separate steps. 
The first step is assembling the detector response function, and the second step is the iterative image deconvolution.

The imaging response function in this case is predominately the \NUSTAR PSF.
However a few factors complicate the image reconstruction process for \NUSTAR. 
First, the PSF is a function of energy and distance from the optical axis --- it elongates farther away from the optical axis, though the fractional change is significantly less than in other focusing instruments such as \CHANDRA due to the larger overall HPD. 
Second, the optical axis is not fixed at one position on the detector, but moves around on the detector during each orbit.
Finally, the PSF is slightly different for each module.
Consequently, a source response function has to be calculated for each observation, module, source, and energy bin, taking into account the movement of the optical axis.

Besides the source distribution, this response function is the key input into the iterative deconvolution algorithm. 
Two approaches have been tested here, the standard Richardson-Lucy algorithm \citep{Richardson1972,Lucy1974}, and a maximum entropy approach \citep{Hollis1992}.
For both methods we performed the deconvolution once with background estimation and once without. 
However, in summary, the approaches show only small differences in the final result after typically 100 iterations. 
Therefore we only show the Richardson-Lucy deconvolved images without separate background estimation.

\begin{figure*}
\centering
\vspace{1.0cm}
\begin{minipage}{.48\textwidth}
  \centering
  \includegraphics[width=0.9\textwidth]{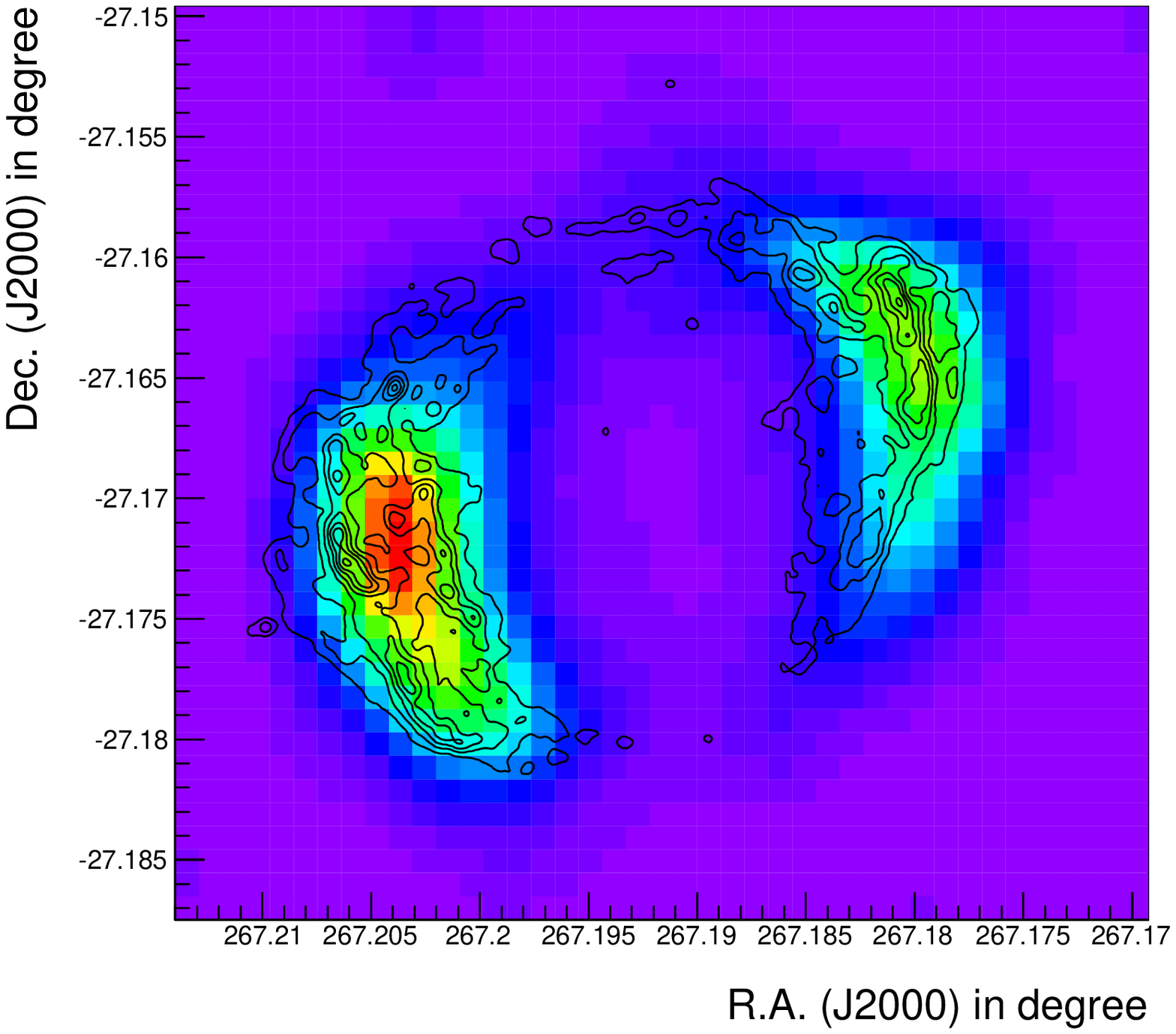}
\end{minipage}%
\begin{minipage}{.48\textwidth}
  \centering
  \includegraphics[width=0.9\textwidth]{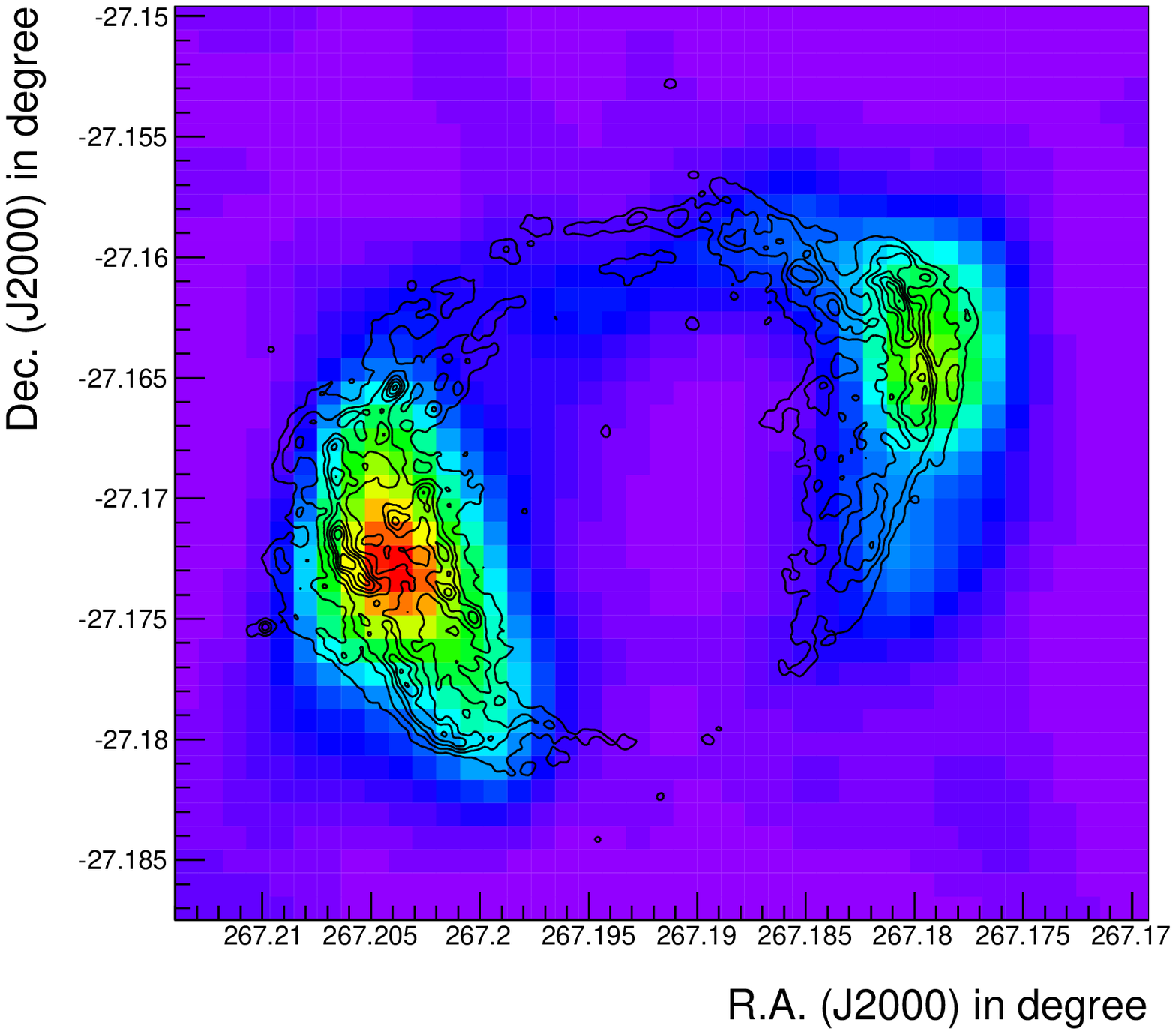}
\end{minipage}
\caption{Deconvolved \NUSTAR image of \G in the energy range from 3-8 keV (left) and 8-20 keV (right). Contours of a \CHANDRA image in the energy range from 3-8 keV are superimposed on both.}
\label{Fig:Deconvolved}
\end{figure*}

Figure \ref{Fig:Deconvolved} shows the deconvolved images of G1.9+0.3 in the energy band from 3-8 keV and 8-20 keV. 
Superimposed is the \CHANDRA image in the energy band from 3-8 keV using observations 12689-12694 (see Table \ref{ChandraObsList}). 
Since the pointing of \NUSTAR is not known to the same accuracy as the pointing of \CHANDRA, the best fit offset between the \CHANDRA image and the \NUSTAR image in the 3-8 keV band has been determined and applied.

\begin{deluxetable}{ccc}
\tabletypesize{\scriptsize}
\tablecaption{List of used \CHANDRA observations of G1.9+0.3\label{ChandraObsList} from 2011}
\tablewidth{0pt}
\tablehead{
\colhead{ID} & \colhead{Date} & \colhead{Exposure [ks]}
}
\startdata
12689 & 14--16 July & 156 \\
12690 & 16--17 May & 48 \\
12691 & 09--11 May & 184 \\
12692 & 12--14 May & 162 \\
12693 & 18--19 May & 127 \\
12694 & 20--22 May & 158 \\
\enddata
\end{deluxetable}

Compared to the undeconvolved images (Figure \ref{Fig:AllBackprojected}), the central region of the SNR is now largely devoid of photons similar to the \CHANDRA image, and the width of the lobes is more closely reproduced. 
The differences between the two images are not significant and mostly due to the deconvolution process itself --- they are of the same magnitude as the differences between the various deconvolution approaches. 

In summary, no significant differences between the emission in the 3 to 8~keV and in the 8 to 20~keV band can be found either in the deconvolved images or in a profile of the remnant to the limits of the data. 
This is a strong indication that the same processes at the same locations are responsible for the generation of the soft and hard X-ray emission.

Finally, our images confirm at higher energies the striking difference in morphology between radio and X-ray images \citep{Reynolds2008b}: the radio maximum is not at either of the NW or SE limbs, but along the bridge of emission that connects them to the north.  Figure 4 shows that the northern bridge is slightly brighter than the southern bridge in X-rays, but is still far fainter than either bright limb. \cite{Reynolds2009} conjecture that the radio peak may not result from shock-accelerated electrons at the blast wave, but at the contact discontinuity between shocked interstellar material and ejecta, which is likely responsible for the bright ring of radio emission interior to the X-ray-defined blast wave in Cassiopeia A \citep{Gotthelf2001}, where turbulent acceleration may be occurring \citep{Cowsik1984}.  Emission from that region would have to have a much lower rolloff frequency so that it does not extend into the \CHANDRA and \NUSTAR bands.  Present radio images do not have sufficient angular resolution to separate a possible blast-wave component from the bright maximum. 

\section{Spectral properties}
\label{sec:spectral}

\subsection{Analysis}

\NUSTAR's excellent high-energy response allows for the first direct measurement of the high-energy tail of the synchrotron X-ray emission of \G.
However, in order to correctly retrieve the low-energy foreground absorption, we still need to perform a combined fit using \CHANDRA and \NUSTAR data. 
Three long \CHANDRA observations (IDs 12691, 12692, 12694) were chosen for the fit, with a combined effective exposure time of 505 ks (see Table \ref{ChandraObsList}). 
While it is generally preferred to use \citet{Wilms2000} abundances in combination with \NUSTAR data, those do not give the best fit for the \CHANDRA data. 
\citet{Reynolds2009} showed that the abundances from \citet{Grevesse1998} reproduce the measured spectra best, and we therefore used them here.
Due to the complicated background conditions, we only fit up to an energy of 30 keV in order to stay in the range where the source photons dominate over the background for the whole SNR. 

\begin{figure*}[ht]
\centering
\includegraphics[angle=270, width=0.8\textwidth]{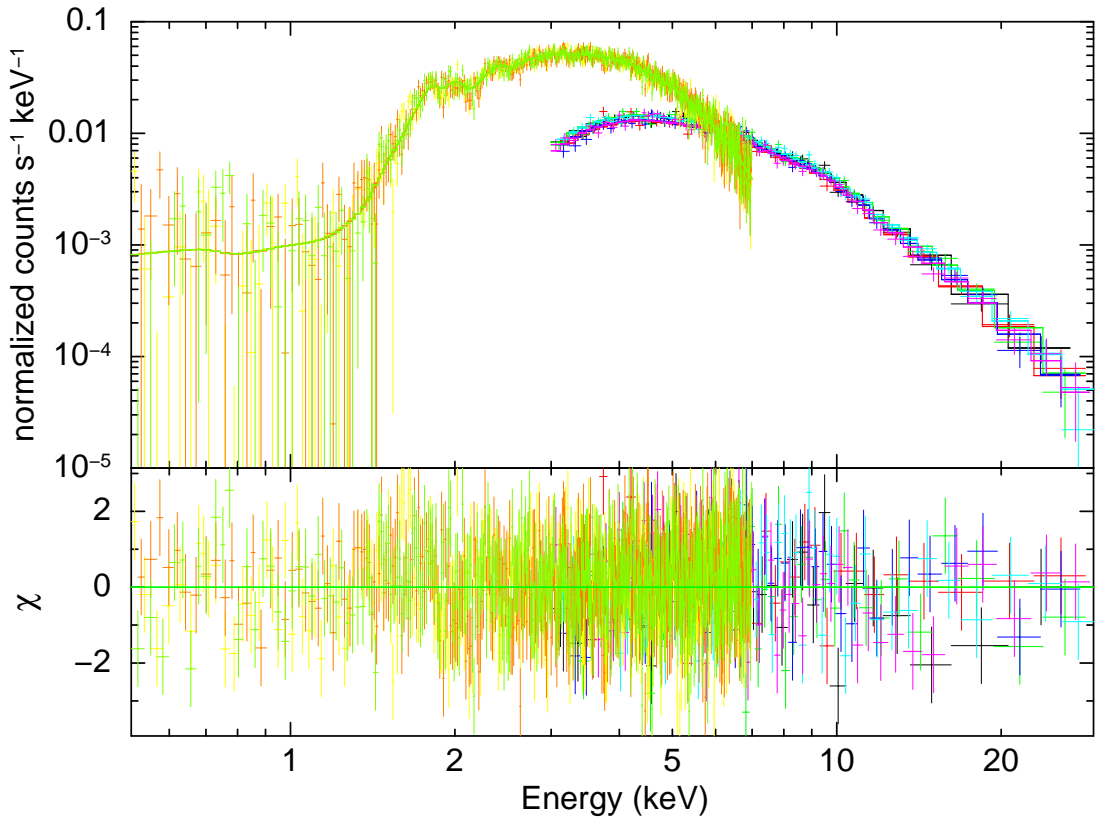}
\caption{Combined fit to the \CHANDRA (low energy range) and \NUSTAR data (high energy range) with the {\tt srcut} model. The differently colored data points and fits correspond to the individual \CHANDRA and \NUSTAR observations (see text and Table \ref{sec:obs}). 
The Figure shows the measured instrument-dependent count rates and not the deconvolved spectra, so no overlap between \CHANDRA and \NUSTAR data is expected.
The radio flux is fixed at the values from \cite{Reynolds2009}. The fit parameters are in agreement with the Chandra results in that paper, albeit with much smaller uncertainties.
\label{Fig:Spectrum_srcut}}
\end{figure*}

The very high column density means that in the \CHANDRA band dust scatters photons from bright source regions into fainter ones, and even beyond the source boundaries altogether.  
These effects were accounted for in \cite{Reynolds2009} in an approximate way, resulting in changes to derived quantities of up to 50\% in \NH and $\nu_{\rm rolloff}$ and 0.15 in $\alpha$.  
While dust scattering should not be important above about 8 keV, the joint fitting includes systematic effects at that level discussed in \cite{Reynolds2009}.

The first task is to reproduce and possibly extend the \cite{Reynolds2009} \CHANDRA results (those ignoring the dust scattering effect) by fitting the \NUSTAR and \CHANDRA data with the exponentially cut-off synchrotron model {\tt srcut} \citep{Reynolds1999}. 
This model describes the emission as originating from a single power-law distribution of electrons with an exponential cutoff.
A required normalization parameter for this model is the radio flux at 1~GHz. 
We use a value of 1.24~Jy, which has been derived by starting with the value from \cite{Reynolds2009}, 1.17~GHz, and accounting for
for the observed increase of the radio flux of roughly 1.2\% per year \citep{Murphy2008}. 
Figure \ref{Fig:Spectrum_srcut} shows that the {\tt srcut} model gives an excellent fit (reduced \chis of 1.06) over the whole energy band from 0.5 to 30 keV.
The derived absorption, \NH~=~$7.23^{+0.07}_{-0.07} \times 10^{22}$\cmnt is slightly above the \cite{Reynolds2009} value (\NH~=~$6.76^{+0.40}_{-0.39} \times 10^{22}$\cmnt), the spectral index, $\alpha=0.633^{+0.002}_{-0.003}$ (vs. $\alpha=0.649^{+0.024}_{-0.024}$) and the roll-over frequency of $\nu_{\rm rolloff}=3.07^{+0.19}_{-0.17} \times 10^{17}$~Hz (vs. $\nu_{\rm rolloff}=5.4^{+4.8}_{-2.4} \times 10^{17}$~Hz) are at the lower end of the Reynolds et al.~(2009) measurements, but are now all much better constrained using the combined data.
Fixing the absorption to the value determined from the combined fit, and then performing the same fit only with the \NUSTAR data basically yields the same fit results (see Table \ref{Models}).

The next step is to determine how much spectral steepening is required in the combined \CHANDRA/\NUSTAR energy band.
Fitting the combined spectrum with a power law (and fixing the \NH to that determined with the {\tt srcut} model) gives a worse fit (reduced \chis of 1.29) with a photon index of $\Gamma=2.52^{+0.02}_{-0.01}$.
Repeating this with a broken power law yields again a good fit with reduced \chis similar to the {\tt srcut} model (1.06), a break energy of E$_{\rm break}=6.5^{+0.5}_{-0.3}$~keV and photon indices of $\Gamma_1=2.40^{+0.02}_{-0.02}$ (low) and $\Gamma_2=2.85^{+0.05}_{-0.04}$ (high).
The {\tt srcut} model gives a power-law spectrum with a cutoff that is roughly exponential in the square root of photon energy.
However, as described in the introduction, there is some motivation for considering a steeper cutoff, one exponential in the photon energy.
Such a model (power-law with exponential cutoff) also describes the observation well, with a reduced $\chi^2$ of 1.07, but a much higher cutoff energy, E$_{\rm cutoff}=15.7^{+1.7}_{-1.4}$~keV, steepening from a power law with photon index $\Gamma = 2.18 \pm 0.04$.  
This indicates that some steepening of the spectrum is definitely necessary at higher energies, but the current data alone cannot determine the best model.

\begin{deluxetable*}{ccccc}
\tabletypesize{\scriptsize}
\tablecaption{Spectral fit models\label{Models}}
\tablewidth{0pt}
\tablehead{ \colhead{Region} & \colhead{Model} & \colhead{Energies} & \colhead{Parameters}  & \colhead{Red. \chis} \\
                             &                 & \colhead{(keV)}          & \colhead{(E in keV, $\nu_{\rm rolloff}$ in $10^{17}$ Hz)}  &  }
\startdata
\CHANDRA+\NUSTAR: Whole & srcut & 0.5-30 & $\alpha=0.633^{+0.002}_{-0.003}$, $\nu_{\rm rolloff}=3.07^{+0.19}_{-0.17}$ & 1.06 \\
\CHANDRA+\NUSTAR: Whole & power law & 0.5-30 & $\Gamma=2.52^{+0.02}_{-0.01}$  & 1.29 \\
\CHANDRA+\NUSTAR: Whole & broken power law & 0.5-30 & $\Gamma_1=2.40^{+0.02}_{-0.02}$, E$_{\rm break}=6.5^{+0.5}_{-0.3}$, $\Gamma_2=2.85^{+0.05}_{-0.04}$ & 1.06 \\
\CHANDRA+\NUSTAR: Whole & exp. cutoff power law & 0.5-30 & $\Gamma=2.18^{+0.04}_{-0.04}$, E$_\mathrm{cut} = 15.7^{+1.7}_{-1.4}$ & 1.07 \\

\NUSTAR: Whole & srcut & 3-30 & $\alpha=0.632^{+0.003}_{-0.003}$, $\nu_{\rm rolloff}=2.97^{+0.19}_{-0.18}$ & 1.10 \\
\NUSTAR: Whole & power law & 3-30 & $\Gamma=2.66^{+0.02}_{-0.03}$ & 1.43 \\
\NUSTAR: Whole & exp. cutoff power law & 3-30 & $\Gamma=2.18^{+0.09}_{-0.09}$, E$_\mathrm{cut} = 15.7^{+3.4}_{-2.4}$ & 1.10 \\

\NUSTAR: North-West & srcut & 3-30 & $\alpha=0.603^{+0.007}_{-0.008}$, $\nu_{\rm rolloff}=2.7^{+0.5}_{-0.5}$ &  0.98 \\ 
\NUSTAR: South-East & srcut & 3-30 & $\alpha=0.647^{+0.005}_{-0.005}$, $\nu_{\rm rolloff}=3.6^{+0.4}_{-0.4}$ &  1.04 \\

\NUSTAR: North-West & power law & 3-30 & $\Gamma=2.66^{+0.06}_{-0.06}$ &  1.74 \\ 
\NUSTAR: South-East & power law & 3-30 & $\Gamma=2.59^{+0.04}_{-0.03}$ &  1.40 \\

\NUSTAR: North-West & exp. cutoff power law & 3-30 & $\Gamma=2.07^{+0.05}_{-0.06}$, E$_\mathrm{cut} = 13.9^{+3.1}_{-2.2}$ &  0.98 \\ 
\NUSTAR: South-East & exp. cutoff power law & 3-30 & $\Gamma=2.09^{+0.03}_{-0.04}$, E$_\mathrm{cut} = 14.9^{+2.1}_{-1.6}$ &  1.04 \\
\enddata
\tablecomments{For all fits a \NH=$7.23\times10^{22}$\cmnt was used. For the {\tt srcut} model a fixed flux of 1.24~Jy at 1~GHz was assumed for the whole remnant, 0.43 Jy for the south-east, and 0.12 Jy for the north-west region. All $\nu_{\rm rolloff}$ values are given in $10^{17}$~Hz. }
\end{deluxetable*}

Table \ref{Models} also presents a summary of the fits of the {\tt srcut} and power-law models to the whole remnant and to the North-West and South-East sections individually using only NuSTAR data.
Our separate {\tt srcut} fits for the two limbs produce values of $\alpha$ that are formally significantly different, while those of $\nu_{\rm rolloff}$ are consistent.  
However, those two parameters are strongly anti-correlated in fits: a steeper (larger) $\alpha$ can be partly counteracted by a higher $\nu_{\rm rolloff}$ in fitting a given spectrum, and those parameters do co-vary in this way between the two regions.  
For this reason, we do not believe there is a significant difference in the spectrum between the two limbs in this energy range.  
This conclusion is supported by the consistent results for the power law and the exponential cutoff power-law fits.

\subsection{Discussion}

The broadband 0.5 -- 30 keV X-ray spectral data require some steepening of the spectrum, since a single power-law fit is inferior to three different parameterizations of steepening: a broken power-law, the {\tt srcut} model, and a power-law with exponential cutoff.  
While the data cannot discriminate among these, the {\tt srcut} model has the best physical justification.  
The broken power-law slopes or break energy result from a purely phenomenological model.  
For the exponentially cut off power-law, there is no obvious physical interpretation for the value of $\Gamma$.  
While the value is not far from the radio energy index steepened by 0.5 (or photon index $\Gamma = 0.63 + 1.5$) as would be expected for radiative energy losses in a homogeneous, time-stationary synchrotron source with continuous injection of a power-law distribution to very high energy, none of those conditions is likely to be the case here. 
Furthermore, in that case one would not expect an additional exponential cutoff.

For the {\tt srcut} model, the maximum electron energy is related to
the roll-off photon energy by E$_{\rm max} = 120 \,(h\nu_{\rm rolloff}/1
\ {\rm keV})^{1/2} (\rm B / \mu{\rm G})^{-1/2}$~TeV \citep[][including
  correction of a numerical error of a factor of 1.9 in the definition
  of $\nu_{\rm rolloff}$]{Reynolds1999}, so the 1.3 keV roll-off energy
we measure implies E$_{\rm max} = 140 \,(\rm B / \mu{\rm G})^{-1/2}$ TeV.
Simple estimates from \cite{Reynolds2008a} assuming Bohm diffusion ($\eta R_J \sim 1$) give E$_{\rm max}({\rm loss}) \sim 100\, (\rm B / \mu{\rm G})^{-1/2} u_8$~TeV for electron acceleration limited by radiative losses. 
Here $u_8$ is the shock velocity in units of $10^8\,{\rm cm}\,{\rm s^{-1}}$.
The value for age-limited acceleration is E$_{\rm max}({\rm age}) \sim 2 \times 10^{-11} (\rm B / \mu{\rm G}) \,u_8^2\, t$ TeV.  
Assuming $u_8 = 14$ and an age of about 100~y, these become E$_{\rm max}({\rm loss}) \sim 1000 \,(\rm B / \mu{\rm G})^{-1/2}$ TeV and E$_{\rm max}({\rm age}) \sim 10 \,(\rm B / \mu{\rm G})$ TeV, respectively.  
The operative process for electrons is the one predicting the lower value of E$_{\rm max}$.  
So a relatively modest magnetic field strength of order 10~$\mu$G is adequate to allow age-limited acceleration to the energies required by the {\tt srcut} model.  
This is close to the lower limit on the interior magnetic field of $11 \ \mu$G which has been derived from the upper limit on the TeV flux determined from H.E.S.S. observations~\citep{ambrowski14}.

If B~$\gapprox 20\ \mu$G, then the loss-limited maximum energy is
lower than the age-limited maximum energy, and would therefore produce
the electron cutoff.  The peak photon energy produced by electrons
with that cutoff energy is, however, independent of the magnetic-field
strength (that is, E$_{\rm max}$ deduced from an observed roll-off has
the same magnetic-field dependence as E$_{\rm max}$ predicted for
loss-limited acceleration).  
Independent of the particular {\tt srcut}
model, electrons with E$_{\rm max} \sim 1000 (\rm B / \mu{\rm G})^{-1/2}$
TeV radiate the peak of their synchrotron spectrum at $h \nu_{\rm max} = 0.193
(E_{\rm max}/100\ {\rm TeV})^2 (\rm B / \mu{\rm G})$ keV, i.e., $\sim 20$
keV here.  Therefore, the photon cutoff energy we obtain from our
power-law with exponential cutoff is not too high to be attained in
\G\ --- although it would not be predicted to cause a cutoff from a
power-law of $\Gamma = 2.18$.  The lack of physical motivation for
this picture makes it considerably less plausible than the equally
well-fitting, but self-consistent, {\tt srcut} picture.

An important consequence of acceleration in \G that is limited by age
to about 100 TeV would be that ions as well would be limited to that
energy.  If the cutoff were due to radiative losses of electrons, ion
acceleration would remain age-limited and could continue on up to much
larger values for larger magnetic-field strengths.

Our detection of curvature in the spectrum at the level required by a
single {\tt srcut} component has an important consequence for particle
acceleration in \G.  
It argues against \G containing a superposition of emission with a broad range of maximum photon energies extending well above the 1.3 keV given by our single {\tt srcut} fit.
Therefore, this value can be taken as characteristic of the bulk of the electron acceleration in \G, and can therefore be used to constrain models for shock acceleration and radiative losses.  
This is in contrast to Cas A, whose integrated X-ray spectrum appears to be a straight power-law from 21 to 120 keV, based on observations with \textit{CGRO}, \textit{BeppoSAX}, and \textit{INTEGRAL} \citep{Renaud2006b}.  
This may point to a fundamental difference in the nature of particle acceleration in \G (probably a Type Ia remnant, encountering more-or-less uniform ISM) and Cas A (a Type IIb remnant, encountering stellar-wind material).  
While non-thermal bremsstrahlung has been proposed for the hard continuum in Cas A \citep{Laming2001}, \NUSTAR observations \citep{Grefenstette2014b} show that the morphology of the hardest X-ray emission is strikingly different from that at lower photon energies, and the site of energization of the required suprathermal electrons could be in weak interior shocks \citep{Laming2001}, though this explanation is still not favored.  
\cite{Vink2008} has shown that for a strong blast wave, rapid Coulomb losses on slightly suprathermal electrons should produce a spectral dip which we do not see in G1.9+0.3, where the highest-energy non-thermal X-rays have the same morphology as at lower energies, and are probably due principally to the forward shock.

\section{Upper limits on \Ti emission}
\label{sec:ti}

\begin{figure*}[ht!]
\centering
\includegraphics[width=\columnwidth, width=0.8\textwidth]{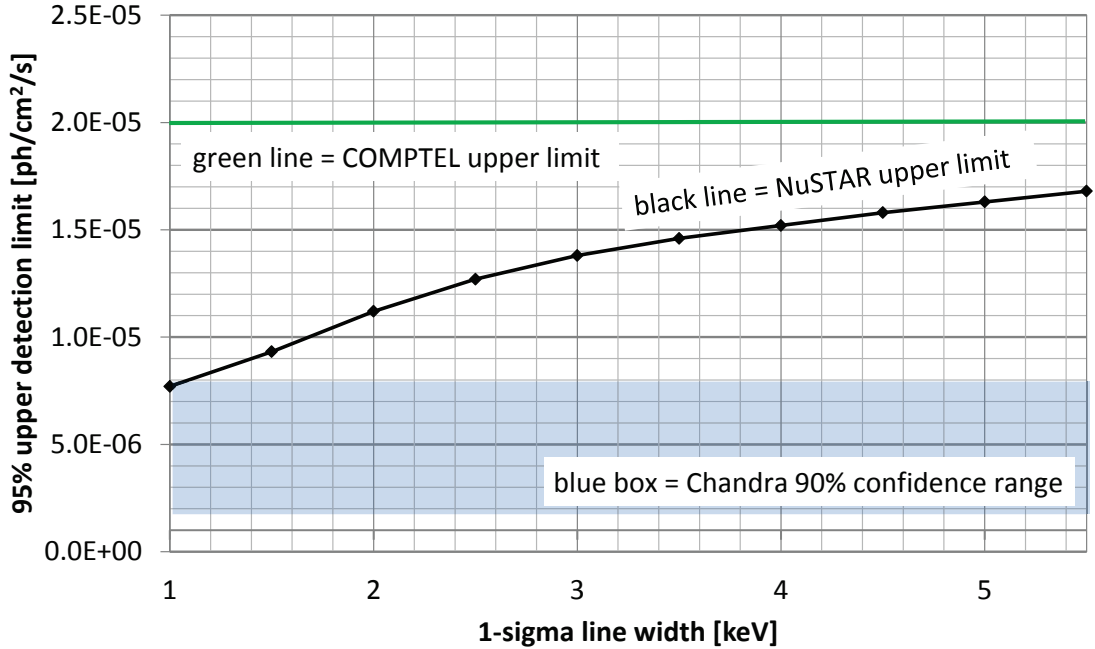}
\caption{95\% upper detection limit for the 67.9~keV line flux from \Ti decay as a function of the line width including the COMPTEL upper limit and the \CHANDRA expectations. See text for details.
\label{Fig:Limit}}
\end{figure*}

Determining the yield and distribution of \Ti in a supernova remnant
is a key tool to understanding supernovae and probing their explosion
mechanism as the majority of the detectable \Ti is produced shortly
before and during the explosion close to the mass cutoff.  
However, for the time being, Cas A remains the only Galactic source for which the \Ti emission has been indisputably measured \citep{Iyudin1994}. 
Recently, \NUSTAR for the first time was able to resolve
the distribution of the \Ti within the Cas A supernova remnant
\citep{Grefenstette2014a}.  
SN1987A, in the Large Magellanic Cloud, also has a robust \Ti detection \citep{Grebenev2012}.  
A third potential source, Vela junior, has a marginal detection by \COMPTEL in the 1.157 keV
line \citep{Iyudin1998}, but was not detected by \INTEGRAL in the 68 and 78~keV lines despite extensive searches \citep{Renaud2006a}.  
\citet{Borkowski2010} recently reported the detection of a 4.1~keV line in the \G SNR with a flux of $1.2^{+1.2}_{-0.85} \times 10^{-6}$ \Flux attributed to the decay chain of \Ti: 
\Ti decays via electron capture to \Sc which, with
a probability of 0.172, yields a 4.1~keV florescence photon to fill
the K-shell vacancy.  Considering a 33\% chance of absorption and
scattering of the 4.1~keV line, a total flux estimate of \Ti in
the 68~keV line of $1.1^{+1.0}_{-0.8} \times 10^{-5}$ \Flux can be
derived.  A reanalysis of the data with more statistics
\citep{Borkowski2013b} reduced the flux estimate for the 68~keV
line to $4.7^{+3.3}_{-3.0} \times 10^{-6}$ \Flux with a width of $4.0^{+1.5}_{-2.9}$~keV (90\% confidence interval) with a corresponding line
width of $35^{+23}_{-24} \times 10^{3}$~km~s$^{-1}$.

In the \NUSTAR energy range, the 67.9~keV line should be the most
easily detectable \Ti line: the 78.4~keV line is very close to the
upper energy limit of \NUSTAR, and the 4.1~keV is harder to detect due
to its lower flux, the higher continuum flux of the remnant itself,
and the significant broadening of the line determined by \CHANDRA.
However, the observed \NUSTAR spectrum of the remnant does not show
any visual evidence of a line at 67.9~keV.  In addition, fitting the
67.9~keV and the 78.4~keV line using a free search for peak center and
peak width does not result in a reasonable fit result.  Therefore, we
derive upper limits by first fitting Gaussian-shaped lines with fixed
peaks at 67.9~keV and the 78.4~keV to the observed spectrum and then
by determining the 95\% upper flux limit using xspec with the same
background extraction regions as before.  Given the
\citet{Borkowski2013b} widths of the 4.1~keV line,
the fit is repeated with different line widths in the range between
1.0 and 5.5~keV (1-sigma width of Gaussian) for the 67.9~keV line.

Figure \ref{Fig:Limit} shows the 95\% upper detection limit for the
67.9~keV line from \Ti decay for the full remnant as a function of the
assumed line width for the full supernova remnant.  The figure also
shows the detection limit as determined with \COMPTEL
\citep{Dupraz1997} for the 1.157~MeV line and with \CHANDRA for the \Sc decay \citep{Borkowski2013b}.  While some of the \NUSTAR
fits also show lower detection limits for the fits with large line
widths, those are very likely just background fluctuations and
therefore ignored here, since similar fits with, e.g., a single 57~keV
line show similar lower detection limits.
As expected, narrower line widths suffer less background and therefore provide stronger limits (see Figure \ref{Fig:Limit}). 

The limits are not affected by stray light on the detector or by
\G itself, since they are undetectable at the energies of the \Ti
lines.  A similar search was performed by narrowing the source
extraction region, but also without success.  In addition, offsetting
the line centers by, e.g., $\pm$5000~km~s$^{-1}$ did not change the results
significantly.

With the current data, \NUSTAR is not yet able to confirm the \CHANDRA
estimate with a direct \Ti line detection.  Determining the yield of
\Ti\ would require significantly longer observation times. For
example, to reach the upper limit of the 90\% confidence range
determined with \CHANDRA for the \Ti flux assuming a 4~keV line width
would require at least 1.4~Ms observation time with \NUSTAR.

\section{Conclusions}

\NUSTAR reproduces the \CHANDRA results concerning morphology and
spectrum of \G, but could not directly detect \Ti emission.  The
morphology of the supernova remnant does not vary significantly
between 3 and 20\,keV. After deconvolution, the \NUSTAR morphology
agrees well with archived \CHANDRA observations.  The data require a
steeping of the spectrum in the combined \CHANDRA/\NUSTAR energy
band. A {\tt srcut} model can describe the spectrum of G1.9+0.3 from
0.5 to 30 keV very accurately.
The fitted roll-off energy of 1.3 keV could result from electron
acceleration limited by the remnant's age of about 100 yr, if the
magnetic field is below about 20 $\mu$G, in which case both electron
and ion spectra would cut off around 100 TeV. The 95\% upper detection
limit for the 67.9 keV line from \Ti decay is roughly $1.5 \times
10^{-5}$ \Flux for an assumed line width of 4.0 keV (1 sigma).  For
the future, at least 4 times longer exposure with \NUSTAR would be
required in order to confirm the \CHANDRA estimate for the \Ti flux.

\acknowledgments

This work was supported under NASA Contract No. NNG08FD60C, and made
use of data from the \NUSTAR mission, a project led by the California
Institute of Technology, managed by the Jet Propulsion Laboratory, and
funded by the National Aeronautics and Space Administration.  We thank
the \NUSTAR Operations, Software and Calibration teams for support
with the execution and analysis of these observations.  This research
has made use of the \NUSTAR Data Analysis Software (\NUSTARDAS)
jointly developed by the ASI Science Data Center (ASDC, Italy) and the
California Institute of Technology (Caltech, USA).
    


{\it Facilities:} \facility{\NUSTAR}, \facility{\CHANDRA}







\end{document}